\documentclass[10pt,journal,compsoc]{IEEEtran}
\usepackage{booktabs} 
\usepackage{graphicx}
\graphicspath{ {./images/} }
\usepackage[labelfont=bf,textfont=it,justification=centering]{caption}
\usepackage{subcaption}
\usepackage{hyperref}
\usepackage{url}

\ifCLASSOPTIONcompsoc
  \usepackage[nocompress]{cite}
\else
  \usepackage{cite}
\fi
\ifCLASSINFOpdf
\else
\fi

\begin{document}
\onecolumn

\title{Music Genre Classification with ResNet and Bi-GRU Using Visual Spectrograms}

\author{Junfei Zhang\IEEEcompsocitemizethanks{\IEEEcompsocthanksitem The University of Melbourne\\
E-mail: junfei.zhang@student.unimelb.edu.au}}

\markboth{2023 Semester 1 - SCIE30001 - Final Report}{}

\IEEEtitleabstractindextext{%
\begin{abstract}
Music recommendation systems have emerged as a vital component to enhance user experience and satisfaction for the music streaming services, which dominates music consumption. The key challenge in improving these recommender systems lies in comprehending the complexity of music data, specifically for the underpinning music genre classification. The limitations of manual genre classification have highlighted the need for a more advanced system, namely the Automatic Music Genre Classification (AMGC) system. While traditional machine learning techniques have shown potential in genre classification, they heavily rely on manually engineered features and feature selection, failing to capture the full complexity of music data. On the other hand, deep learning classification architectures like the traditional Convolutional Neural Networks (CNN) are effective in capturing the spatial hierarchies but struggle to capture the temporal dynamics inherent in music data. To address these challenges, this study proposes a novel approach using visual spectrograms as input, and propose a hybrid model that combines the strength of the Residual neural Network (ResNet) and the Gated Recurrent Unit (GRU). This model is designed to provide a more comprehensive analysis of music data, offering the potential to improve the music recommender systems through achieving a more comprehensive analysis of music data and hence potentially more accurate genre classification.

\end{abstract}

}

\maketitle
\IEEEdisplaynontitleabstractindextext
\IEEEpeerreviewmaketitle
\ifCLASSOPTIONcompsoc
\else
\fi

\section{Introduction}

The landscape of music consumption has significantly shifted in recent years with the advent of music streaming services \cite{melspec}, such as Spotify, Apple Music, Amazon Music and YouTube Music. As a result, music recommendation systems have become an essential part of these platforms to enhance user experience and satisfaction, which is seen to have a significant entertainment and commercial value. Despite these substantial advancements, music recommendation systems often struggle with the inherently complexity of music data \cite{melspec}. This complexity stems largely from the challenges involved in comprehending and classifying the intricacies embedded within musical genres \cite{comparison}. Current systems heavily rely on user input and manual genre classification, often leading to recommendations that fail to accurately capture the listener's preferences \cite{hacker}. Furthermore, with the vast number of music files currently exist and continuously emerging on the internet, manual classification becomes increasingly unsustainable, leaving numerous songs unclassified or inaccurately labeled, as the industrial standard currently is incomplete \cite{carlos}. These challenges underline the critical need for a more advanced, automatic system for music genre classification, necessitating a shift towards Music Information Retrieval (MIR)-based solutions. An Automatic Music Genre Classification (AMGC) system becomes imperative. Such a system could autonomously analyze and classify music into genres, considering a broader range musical features and their complexities. This would not only streamline the classification process, but also greatly improve the quality and personalisation of recommendations.

Music genre classification presents a significant challenge due to the lack of strict boundaries between genres \cite{george}, which is compounded by the complexity of music's interaction with the public and the unique vibe it carries \cite{george}. In the realm of AMGC, traditional machine learning techniques, such as k-Nearest Neighbors (k-NN), Support Vector Machines (SVM), and Random Forest (RF) have demonstrated considerable potential. However, these algorithms heavily rely on the effectiveness of manually engineered features and feature sets, which are typically defined by the music's instrumentation (eg presence of electric guitars), rhythmic structure (eg beat), and harmonic content (eg chords, tones) \cite{george}. The reliance on the analysis of feature importance is a significant limitation as it fails to fully capture the complexity of music data. To address this limitation, deep learning methods such as Convolutional Neural Networks (CNN) have shown significant promise in handling this task \cite{audio}. While they also often operate on engineered features, such as the Short-time Fourier
transform (STFT)\cite{featurerepresentation} and the Mel Spectrograms \cite{melscale}, their ability to learn hierearchical representations allows them to extract complex patterns from this transformed representation of the audio, potentially improving the genre classification performance. Traditional CNNs are primarily designed to deal with spatial or hierarchical data mostly present in image data, whereas it does not inherently model the sequence of data with temporal dynamic features \cite{bottomup}, thereby possibly limiting their effectiveness for time-series data commonly used to represent audio. Given the shortcomings of traditional machine learning and acknowledging the limitations and potential of deep learning methods, there is a strong motivation to design a more effective approach. This study aims to employ deep learning to manage the complexity of the task and propose a deep learning architecture that can effectively extract both hierarchical and temporal information from music data effectively. This paper presents an innovative approach of using visual spectrograms as image-based inputs, and propose a novel hybrid model that combines the strength of the Residual Neural Network (ResNet) and Gated Recurrent Unit (GRU). This hybrid model processes image-based input with ResNet while the GRU block handles sequence-like data from the reshaped image inputs, effectively leveraging the strength of both architectures to provide a more comprehensive analysis of the music data. The implications of this research could significantly improve the music recommendation systems by offering a more precise genre classification, thereby potentially enhancing user experience on music streaming platforms.

This paper outlines the development and evaluation of the proposed architecture. The organisation of the paper is as follows: Section 2 gives a review of current research related to the task. Section 3 describes the dataset used for training and testing the model, as well as the data augmentation. Section 4 details the methodology and the classification pipeline. Section 5 analysis the results obtained, while Section 6 discusses the implications of the study. Section 7 summarises the conclusions drawn from the research, as well as a suggestion of potential future work. Appendix A showcases a deployment of the classifier to a music recommendation system.

\section{Related Work}

\subsection{Problem Definition}
Music genre classification is a difficult task in nature due to the ambiguity of genre boundaries. Highlighting this, a research study by Robert et al. \cite{robert} investigated manual music classification among college students. Participants were asked to listen to music pieces and classify them into one of ten genres: blues, country, classical, dance, jazz, latin, pop, R\&B, rap, and rock. The research found that participants were able to achieve only 70\% classification with 3 seconds musical pieces, and the accuracy did not improve for longer pieces. This is always referenced as a benchmark in machine genre classifications that sets the stage of the AMGC task. 

The AMGC problem was initially defined by Tzanetakis and Cook \cite{george} in 2002. In current context, the original sound wave is sampled and transformed into a sequence of numeric values $S=<s_1, s_2 ... s_N>$ \cite{carlos}, where $s_i$ represents the sample at instant $i$, and $N$ represents the total number of samples. The features that contained in the sequence can be extracted to get the feature vector $\bar{X}=<x_1, x_2 ... x_D>$. The AMGC problem concerns about finding the genre from a set of music genres that best represents the genre of the music \cite{carlos}. It can be expressed using a statistical perspective by this equation: \\ 
\begin{equation} \label{eq:1}
    \hat{g} = argmax_{g\in G} P(g|\bar{X})
\end{equation}
where $P(g|\bar{X})$ is the probability that the music belong to the genre $g$ given the features expressed by $\bar{X}$. This equation is applicable to both machine learning and deep learning approaches. This section presents the related works and solutions pertinent to the AMGC task, providing insights on the evolution of this field to its current state. A summary of the performance is presented in Table \ref{tab:accuracy}.

\begin{table}[h!]
    \centering
    \caption{A summary of current methods and accuracy for the AMGC task (10 genres)}
    \label{tab:accuracy}
    \begin{tabular}{@{}lp{5cm}}
        \toprule
        Methods & Accuracy \\
        \midrule
        \multicolumn{2}{l}{\textit{Machine Learning}} \\
        KNN & 0.54\textsuperscript{\cite{stanford}}, 0.61\textsuperscript{\cite{george}}, 0.63\textsuperscript{\cite{turkey}} \\
        SVM & 0.60\textsuperscript{\cite{stanford}}, 0.65 \textsuperscript{\cite{carlos}}, 0.73\textsuperscript{\cite{turkey}} \\
        NB & 0.57\textsuperscript{\cite{turkey}} \\
        DT & 0.55\textsuperscript{\cite{turkey}} \\
        RF & 0.81\textsuperscript{\cite{medium}}, 0.66\textsuperscript{\cite{turkey}} \\
        
        \addlinespace
        \multicolumn{2}{l}{\textit{Neural Networks}} \\
        CNN & 0.69\textsuperscript{\cite{turkey}}, 0.72\textsuperscript{\cite{sugianto}}, 0.82\textsuperscript{\cite{stanford}} \\
        YOLOv4 & 0.945\textsuperscript{\cite{melspec}} \\
        BBNN & 0.939\textsuperscript{\cite{bottomup}} \\

        \bottomrule
    \end{tabular}
\end{table}

\subsection{Machine Learning Approach}
For traditional machine learning methods, Equation \ref{eq:1} models the relationship between input features and output labels. Examples of such techniques include K-Nearest Neighbors (KNN), Support Vector Machine (SVM), Naive Bayes (NB), Decision Tree (DT), and Random Forest (RF). These techniques have shown considerable promise in the domain of music genre classification. For instance, a study \cite{stanford} demonstrated testing accuracy of 0.54 using KNN and 0.60 using SVM. Another comparative evaluation \cite{turkey} reported accuracy scores of 0.63 for KNN, 0.73 for SVM, 0.57 for NB, 0.55 for DT, and 0.66 for RF. However, it's crucial to remember that these performances are largely dependent on the effectiveness of the selected features or feature sets. Thus, the quality and relevance of feature selection is a crucial consideration in traditional machine learning approaches and remains an open question in the Automatic Music Genre Classification (AMGC) problem \cite{carlos}.

Further studies have tried to delve deeper into the significance of different feature sets. The noteworthy study by George et al. \cite{george} explored the KNN model using three sets of features representing timbral texture, rhythmic content, and pitch content. The study found that timbral texture features outperformed other sets, underscoring their relevance in the music classification process. In another work \cite{carlos}, researchers introduced space and time decomposition strategies which used One-Against-All (OAA) and Round-Robin (RR) approaches to obtain the final genre label based on an ensemble method applied to the initial, middle, and end parts of the music. The optimal result, which was an accuracy of 0.65, was achieved with the SVM classifier and space-time decomposition using Round Robin (RR) approach. The study shows that 12 out of the 14 most selected features are timbral texture related, such as 1 st. MFCC mean, 3 rd. MFCC mean, Standard deviation for spectral flow, Standard deviation for 1 st. MFCC, etc. However, the authors noted that different features might have varying importance based on the classifier used (KNN, NB, SVM), as well as different classification approaches on different segments of music. This complexity underlines a fundamental challenge in traditional machine learning for AMGC: despite its promising performance, its effectiveness heavily depends on the quality of feature selection and its ability to capture the underlying structure of music data, motivating the exploration of alternative approaches such as deep learning.

\subsection{Deep Learning Approach}
Deep learning approaches use a layered structure of artificial neurons to model complex patterns in music data. They do not explicitly calculate the probability in Equation \ref{eq:1} but implicitly learn to approximate $P(g|\bar{X})$ through their architecture. The training process adjusts the weights of the connections in the network to minimize a loss function. This can potentially lead to more robust and nuanced feature extraction.

Convolutional Neural Networks (CNN) have been recognized for their superior accuracy in AMGC tasks compared with the machine learning approaches. A comparative study showcased CNN's supremacy over other machine learning models, including Support Vector Machines (SVM), Random Forests, and eXtreme Gradient Boosting (XGB) \cite{mlreview}. While CNNs are traditionally known for their image pattern recognition capabilities, their application to AMGC tasks often involves time-series data as the input since it is a straightforward representation of audio signals. However, current research highlights promising results when using image-based inputs such as the visual spectrograms. Two studies have reported high accuracies when applying CNN to spectrogram images for music genre classification, one achieving an accuracy of 0.874 using 2.56-second slices and 0.918 using full song data \cite{review}, while the other using the CNN based YOLOv4 architecture on visual Mel Spectrograms reported an accuracy as high as 0.945 \cite{melspec}. Using graphical spectrogram representations not only offers high generalization performance but also eliminates the need for constructing a specialized audio model \cite{melspec}. 

There are multiple time-frequency image formation proposed for CNN. For tasks like speech emotion recognition \cite{speech}, the traditional spectrogram representation formed using the short-time Fourier transform (STFT) \cite{frequencyrepresentation} is often employed \cite{audio}. However, this method can be computationally expensive if the goal is to uncover unique frequency characteristics by using a large number of points in the Fourier transform \cite{audio}. As a more efficient alternative, methods have been proposed to utilize a larger transform length and compute the filterbank energies in frequency subbands \cite{mfcc}. The mel-filter is one such frequently used filter in this context, which utilizes the mel-scale \cite{melscale} to space the frequency bands. Image representations employing the mel-filter, such as Mel Spectrograms, have been widely used with CNN in various acoustic classification challenges \cite{audio} \cite{dcase}. Current advancements in the AMGC task leverage deep learning approaches, specifically CNN, for their remarkable accuracy and efficiency. The following summary presents these high performing approaches in detail.

\subsubsection{Bottom-up Broadcast Neural Network (BBNN) \cite{bottomup}}
This study focuses on addressing a key limitation of traditional CNN architectures in the context of music genre classification, which is the inability for CNN to learn adequate and comprehensive features for music genre classification in temporal dynamics. Their innovative CNN structure, Bottom-up Broadcast Neural Network (BBNN), effectively utilizes both high-level semantic information and low-level features from music data. It achieves this by constructing a strategic arrangement of blocks and connections among them. This arrangement is designed to maximize the use of and preserve low-level information as it is transmitted to higher layers of the network. With this design, the proposed BBNN achieved an impressive accuracy of 0.939 on the GTZAN dataset, signifying a significant advance in the AMGC realm.

While the BBNN model represents a significant advancement in automatic music genre classification, there is still room to apply transfer learning and leverage the power of well-established pretrained models. Among them, the Residual Networks (ResNet) and You Only Look Once (YOLO) architectures stand out for their efficiency and depth, facilitating the training of much deeper networks, which could potentially enhance the learning process. 

\subsubsection{You Only Look Once Version 4 (YOLOv4) \cite{melspec}}
The study under discussion introduces an innovative approach to the AMGC problem, employing the YOLOv4 architecture using visual Mel Spectrograms. The promising result of an accuracy of 0.945 signifies the potential of YOLOv4 for such tasks. Originating from the field of computer vision, YOLO (You Only Look Once) models are typically deployed for object detection tasks \cite{yolo}. The fourth iteration of this model series, YOLOv4 \cite{yolov4}, has demonstrated enhancements in speed and accuracy over its predecessors. The architecture of YOLOv4 is composed of five modules based on CNN. Although the conventional application of YOLOv4 is object recognition in images, such as vehicle detection for autonomous driving \cite{autodrive}, or face detection \cite{facedetect} and recognition in surveillance systems \cite{surveillance}, its success in the AMGC task suggests a potential for transferring pre-trained computer vision models to audio classification tasks.

\section{Dataset}
This study utilizes the GTZAN Dataset\cite{gtzanorigin}, which is a highly-regarded public dataset frequently employed for evaluation in machine listening research, specifically for music genre recognition. The dataset sources vary widely, encompassing personal CDs, radio transmissions, and microphone recordings. It is assembled by Tzanetakis and Cook \cite{gtzanorigin} and has served as a significant resource in this research area. The details of the GTZAN Dataset are summarized in Table \ref{tab:data}.

\begin{table}[h!]
    \centering
    \caption{GTZAN Dataset Details}
    \label{tab:data}
    \begin{tabular}{@{}lp{2cm}p{9cm}@{}}
        \toprule
        Item & Data & Details \\
        \midrule
        Genre & 10 & blues, classical, country, disco, hiphop, jazz, metal, pop, reggae, rock \\
        Length & $\sim$30s per clip & Minimum length: 29.93s; Maximum length: 30.65s; Mean length: 30.02s \\
        Sample rate & 22,050 Hz & \\
        Song format & WAV & \\
        Total & 1000 & 100 files/genre \\
        \bottomrule
    \end{tabular}
\end{table}

\subsection{Data Augmentation}
To make the input more robust, this study applies data augmentation to the GTZAN dataset. The original GTZAN dataset comprises 1,000 audio clips of slightly varying lengths, a format not ideally suited for model input. Therefore, to rectify this as well as increasing the number of data, each audio clip in the original dataset was randomly sampled within a contiguous 3-second window at five non-overlapping random locations. This process augmented the dataset to 5,000 audio clips, each lasting precisely three seconds. These details are summarized in Table \ref{tab:augdata}.

\begin{table}[h!]
    \centering
    \caption{Augmented Dataset Details}
    \label{tab:augdata}
    \begin{tabular}{@{}lp{2cm}p{9cm}@{}}
        \toprule
        Item & Data & Details \\
        \midrule
        Genre & 10 & blues, classical, country, disco, hiphop, jazz, metal, pop, reggae, rock \\
        Length & 3s per clip \\
        Sample rate & 22,050 Hz & \\
        Song format & WAV & \\
        Total & 5000 & 500 files/genre \\
        \bottomrule
    \end{tabular}
\end{table}

\section{Method}
\subsection{Preprocessing}
\begin{figure}[h!]
    \centering
    \includegraphics[scale=0.34]{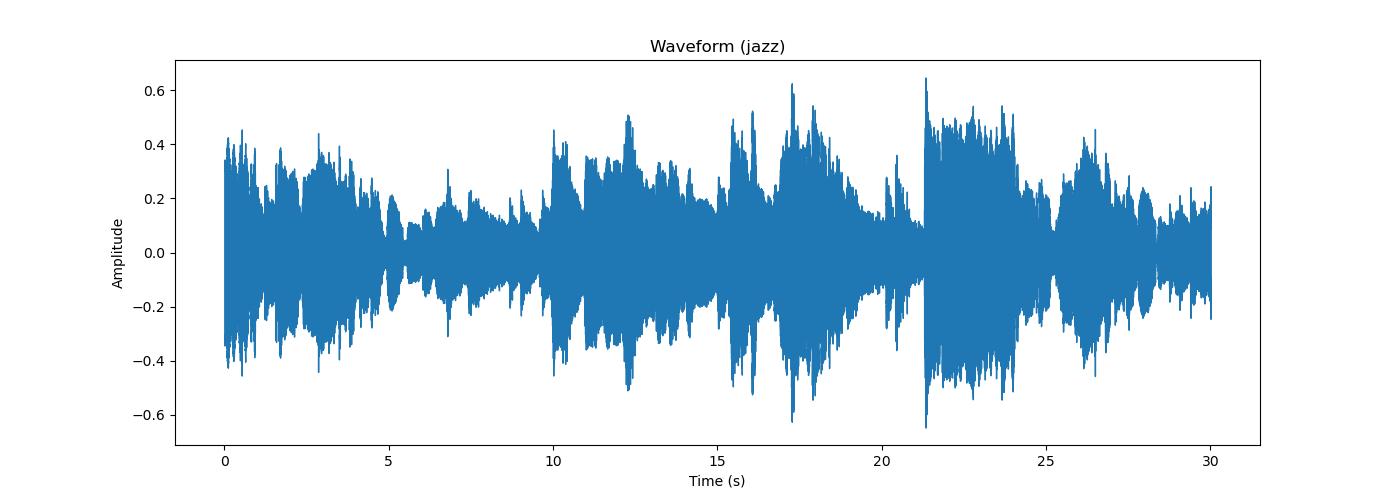}
    \caption{Raw audio waveform for jazz class}
    \label{fig:raw}
\end{figure}
This study utilizes visual Mel Spectrograms as its primary data. Developed by Davis and Mermelstein \cite{melorigin} in 1980, the mel-scale prioritises specific frequency components and filtering less significant ones, which effectively models the logarithmic frequency resolution of human auditory perception. To convert raw audio data (Fig. \ref{fig:raw}) into Mel Spectrograms, the initial step is to shift the audio from the time domain to the frequency domain using the Short Time Fourier Transform (STFT). The $i$-th feature of this transformation is defined by the following equation:

\begin{equation}
    x_i = STFT\{s\}(i,\omega)=\sum_{n} s_n \cdot w[n - i] \cdot e^{-j\omega n}
\end{equation}

\begin{figure}[h!]
    \centering
    \includegraphics[scale=0.4]{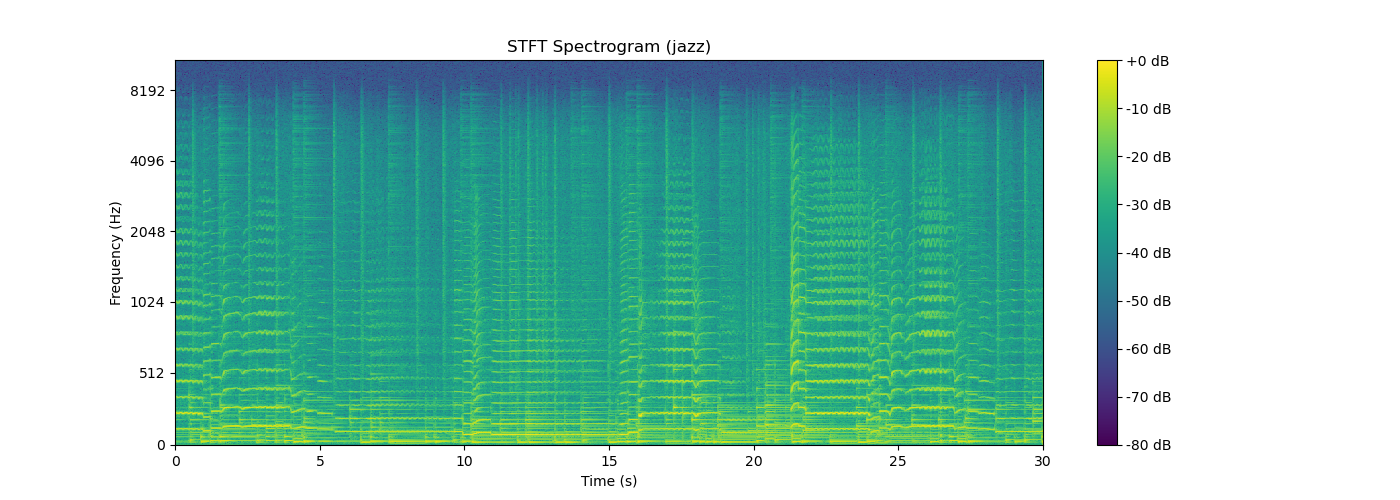}
    \caption{STFT Spectrogram for jazz class}
    \label{fig:stft}
\end{figure}

In this equation, $s_n$ signifies the $n$-th sample in the sequence $S$, $w[n - i]$ represents a window function centered on the $i$-th sample, and $\omega$ denotes frequency. The outcome, $\bar{X}_{stft}$, is a sequence of time-frequency representations derived via the STFT at various moments within the audio signal. The transitioned visual STFT Spectrogram is presented in Fig. \ref{fig:stft}, demonstrating the transformation of the raw audio from the time to time-frequency domain.
The subsequent phase is the computation of $\bar{X}_{mel}$, the Mel Spectrogram. If $\bar{X}
_{stft,i}$ symbolizes the $i$-th STFT feature, the corresponding Mel frequency feature $\bar{X}_{mel,i}$ can be calculated as:

\begin{equation}
    \bar{X}_{mel, i} = M \cdot |\bar{X}_{stft,i}|^2
\end{equation}

\begin{figure}[h!]
    \centering
    \includegraphics[scale=0.34]{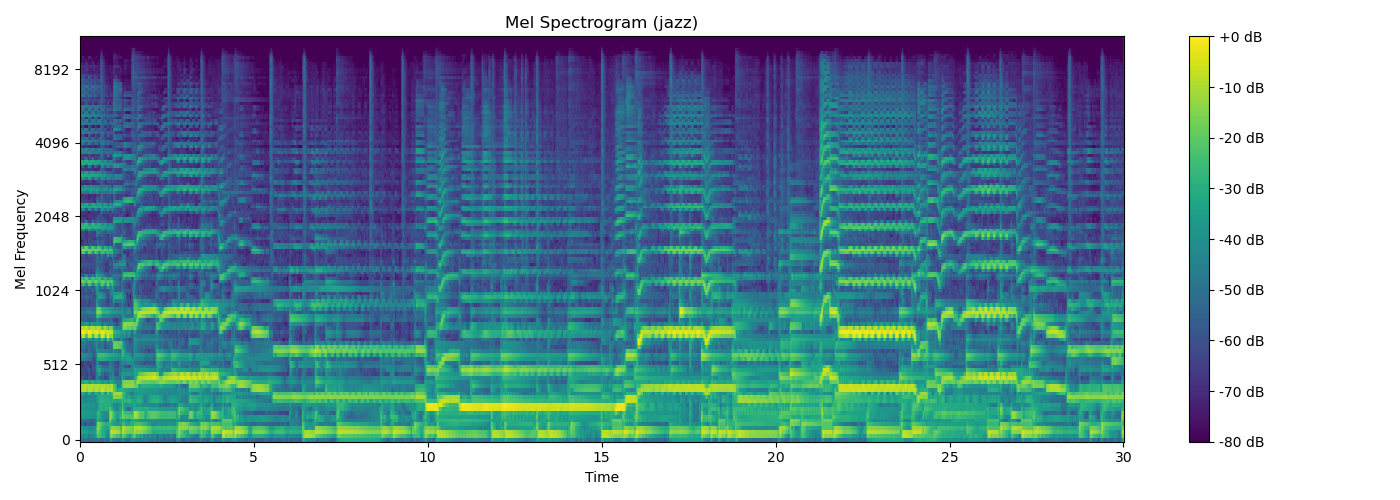}
    \caption{Mel Spectrogram for jazz class}
    \label{fig:mel}
\end{figure}

In this formula, $M$ refers to a matrix depicting the Mel filter bank, and squaring equates to the power spectrum. Each row in $M$ pertains to a unique Mel filter. This matrix is derived based on the mel-scale; the conversion from Hertz ($f_{Hz}$) to Mel ($m$) scale can be described by the following equation:
\begin{equation}
m = 2595 \log_{10}(1 + \frac{f_{Hz}}{700})
\end{equation}
Multiplying this with the power spectrum yields the energy within the Mel-scaled frequency bands. This process is repeated for each STFT feature to attain a corresponding Mel-frequency feature. The logarithm of $\bar{X}_{mel}$ is then computed to acquire the log-Mel-spectrum. Fig. \ref{fig:mel} portrays the Mel Spectrogram. All these operations are implemented via the Librosa library. The parameters are displayed in Table \ref{tab:parameters}.

\begin{table}[h!]
    \centering
    \caption{Parameter Values for Mel Spectrograms}
    \label{tab:parameters}
    \begin{tabular}{@{}lc@{}}
        \toprule
        Parameter & Value\textsuperscript{} \\
        \midrule
        Audio Length (seconds) & 3 \\
        Sampling Rate (hertz) & 22,050 \\
        Window Length (frames) & 2,048 \\
        Overlap Length (frames) & 512 \\
        FFT Length (frames) & 2,048 \\
        Num Bands (filters) & 128 \\
        \bottomrule
    \end{tabular}
\end{table}

\begin{figure}[h!]
    \centering
    \includegraphics[scale=0.55]{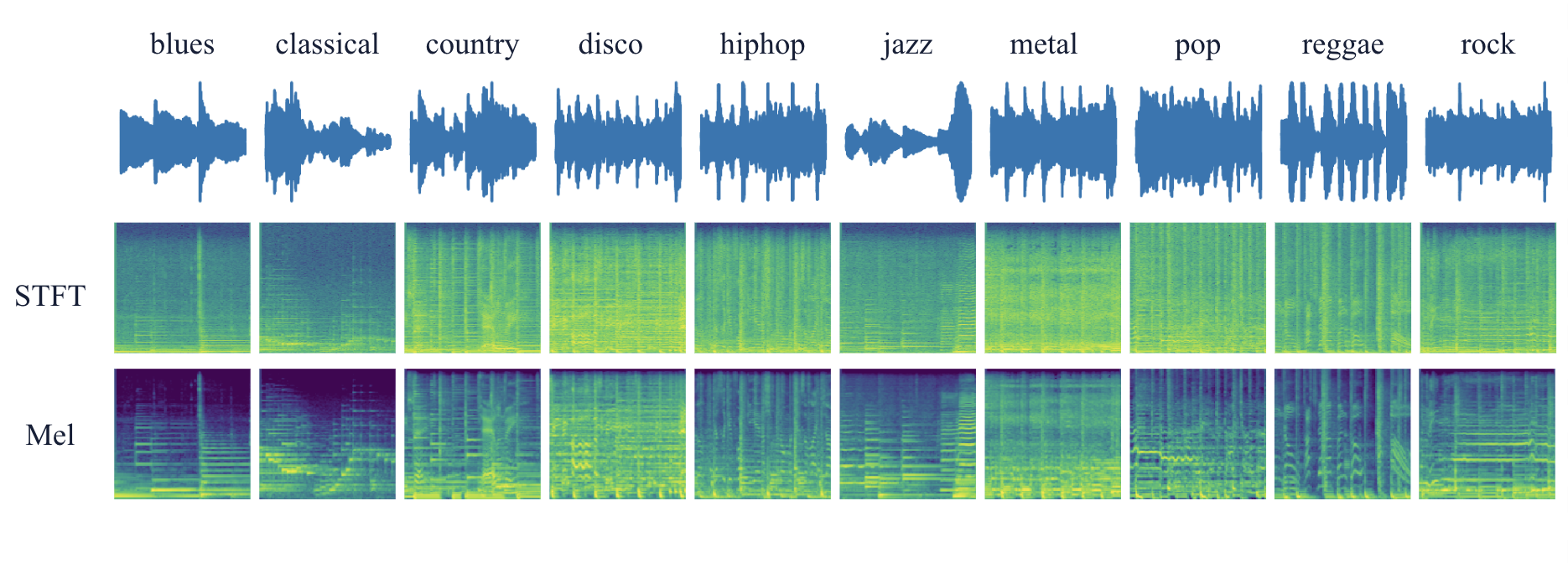}
    \caption{Image inputs}
    \label{fig:input}
\end{figure}

To optimize the image for the architecture, scales and whitespaces are eliminated from the spectrograms, and images are reshaped to the dimensions $(224, 224)$ to be compatible with ResNet input. The finalized input images are presented in Fig \ref{fig:input}.

\subsection{Baseline}
The study starts with implementing two simpler machine learning models as baselines. K-Nearest Neighbors (KNN) and Support Vector Machines (SVM) are being implemented on time-series data as the input. The performance of these two models can be seen via Table \ref{tab:accuracy2}.
\begin{table}[h!]
    \centering
    \caption{Machine Learning Method Accuracy}
    \label{tab:accuracy2}
    \begin{tabular}{@{}lccc@{}}
        \toprule
        Methods & Raw data & STFT Spectrogram & Mel Spectrogram \\
        \midrule
        KNN: & & & \\
        n = 10 & 0.21 & 0.51 & 0.52 \\
        n = 15 & 0.21 & 0.51 & 0.51 \\
        n = 20 & 0.21 & 0.51 & 0.48 \\
        SVM & 0.09 & 0.60 & 0.57 \\
        \bottomrule
    \end{tabular}
\end{table}

Overall, both KNN and SVM perform significantly better on STFT and Mel Spectrograms than on raw data. The performance on STFT and Mel Spectrograms is relatively similar for KNN, whereas for SVM, the STFT Spectrogram yields better accuracy. These results suggest that the spectrogram representations are much more informative for the KNN and SVM models than raw data. However, the KNN approach is not significantly sensitive to the human-aligned features introduced by the mel-scale, and the SVM model works better with STFT Spectrograms.

\subsection{Convolutional Neural Networks (CNNs) \& Resudual Networks (ResNets)}
CNN \cite{imagenet} is the base architecture for ResNet \cite{resnet}, which is used in this study. Conventional CNN consists of convolutional layer, pooling layer, and fully connected layer \cite{melspec}. 

\begin{figure}[h!]
    \centering 
    \includegraphics[scale=0.55]{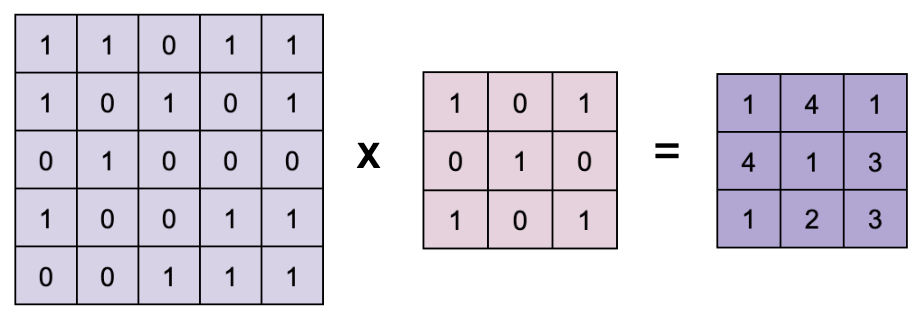}
    \caption{Convolutional Layer}
    \label{fig:conv}
\end{figure}

The Convolutional Layer is the core building block of a CNN. A set of learnable kernels are consisted in the layer's parameters, which is convoluted across the width and height of the input volume during the forward pass. Each kernel is convoluted across the width and height of the input volume, and the dot produce between the entries of the kernel and the input is calculated, producing a 2-dimensional activation map of the kernel. The network learns the kernel that activate when it detects some specific type of feature at some spatial position in the input. The primary purpose is to detect local conjunctions of features from the previous layer. An illustration is shown at Fig. \ref{fig:conv}.

The Pooling Layer (also known as down-sampling) reduces the dimensionality of each feature map while retaining the most important information. Max pooling (Fig. 7) is preferred in this study due to its ability to highlight dominant features. The important rhythmic and harmonic patterns are often distinctive for music genres, which can be preserved by max pooling. 

\begin{figure}[h!]
    \centering 
    \includegraphics[scale=0.45]{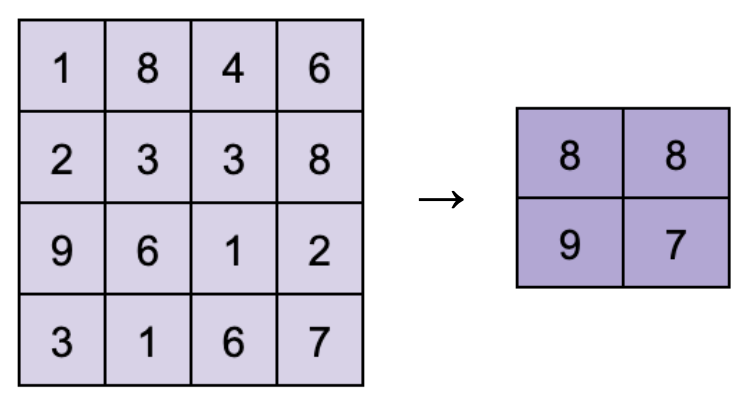}
    \caption{Max Pooling Layer}
    \label{fig:maxpool}
\end{figure}

The Fully Connected layer has its neuron in the previous layer connect to every neuron in the next layer. The purpose of the fully connected layer is to perform classification on the features extracted by the convolutional layers which are downsampled by the pooling layers. 

\begin{figure}[h!]
    \centering 
    \includegraphics[scale=0.45]{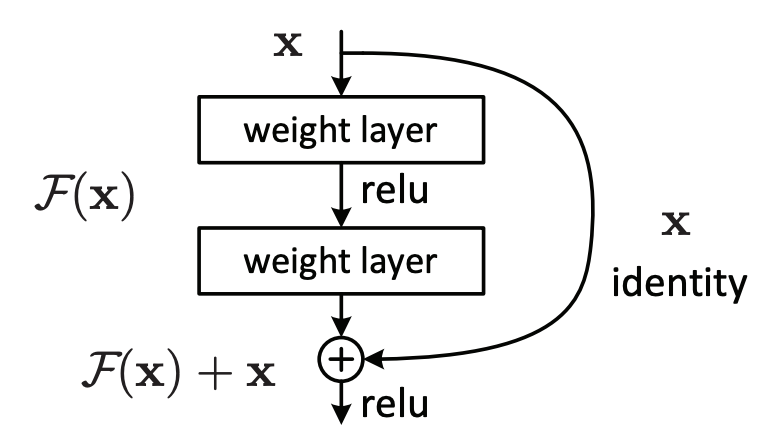}
    \caption{ResNet: Skip Connections \cite{resnet}}
    \label{fig:resnet}
\end{figure}

ResNets \cite{resnet} are a type of CNNs and is used in this study. The core idea behind ResNets is the "skip connections", which allow the gradient to be directly back-propagated to earlier layers. The purpose is the deal with the problem of conventional CNNs; they are hard to train due to the "vanishing gradients" problem. This means that the earlier layers learn very slowly and have difficulty making significant changes, which hinges to train very deep networks. ResNet addresses the problem by adding shortcut connections. This allows the model to carry forward the previously learned information through shortcut connections and directly add it to the later layers. As shown by Fig. \ref{fig:resnet}, instead of trying to learn an underlying function $H(x)$, the stacked nonlinear layers are fitted to another mapping of $F(x) := H(x) - x$. The original function thus becomes $H(x) = F(x) + x$. The ResNet structure makes it possible to train very deep networks which can improve the performance in various computer vision tasks. In this study, ResNet18 is found to be the best performing ResNets. ResNet18 is a member of the Resnet family that has 18 layers with weights. The Layers are structured as follows:
\begin{itemize}
  \item Initial Convolutional Layer (7x7, stride 2)
  \item Max Pooling Layer (3x3, stride 2)
  \item 2 x Convolutional Block (2 layers with 64 filters)
  \item 2 x Convolutional Block (2 layers with 128 filters)
  \item 2 x Convolutional Block (2 layers with 256 filters)
  \item 2 x Convolutional Block (2 layers with 512 filters)
  \item Average Pooling Layer
  \item Fully Connected Layer
\end{itemize}

CNNs are suitable for the AMGC task when using visual spectrograms as inputs due to their capacity to effectively learn hierarchical patterns in image-like data \cite{imagenet}. The spectrograms convert the audio signals into 2-dimensional representations which show how the spectral density of frequencies varies with time, essentially transforming the music data into image-like structures. The hierarchical nature of CNNs makes them able to learn the complex patterns in the spectrograms that can potentially represent basic musical features such as chords and timbre to more intricate structures such as melody and rhythm \cite{bottomup}. However, while conventional CNNs are excellent at capturing spatial hierarchies and patterns, they inherently lack the ability to effectively capture temporal dynamics over time \cite{bottomup}. These temporal dynamics are crucial in identifying musical features such as tempo, rhythm, beat, meter, and the form and structure. This motivates the need of other architecture that are more suitable for processing sequential data.

\subsection{Recurrent Neural Networks (RNN) \& Bidirectional Gated Recurrent Unit (Bi-GRU)}
Recurrent Neural Networks (RNNs) \cite{rnnreview} are a type of neural network that are designed to recognize patterns in sequences of data, which is often being used in speech recognition \cite{speechrecog} and sound event detection \cite{bigrureview}. RNNs make use of a "memory", which performs the same task for every element of a sequence, with the output depended on the previous computations. This is achieved by a looping structure that passes the information from one step of the network to the next step. This can be explained by the mathematical expression:
\begin{equation}
    h_t = \tanh{(W_{hh} \cdot h_{t-1} + W_{xh} \cdot x_t)}
\end{equation}
where $h_t$ is the hidden state at time $t$, $x_t$ is the input at time $t$, and $W_{hh}$ and $W_{xh}$ are weight matrices (learned parameters). However, basic RNNs also suffer from the "vanishing gradient" problem and struggle to learn long-term dependencies due to the decline of gradient during back-propagation, making them hard to carry information through many time steps. As a solution, architectures like Long Short-Term Memory (LSTM)\cite{lstmreview} units and Gated Recurrent Units (GRU)\cite{grureview} were developed, which is able to selectively forget and remember information over long periods. For this study, the GRU unit is employed due to its efficiency over LSTMs. 
The GRU unit is more complex than the conventional RNNs in terms of the hidden state update mechanism. It has two gates: a reset gate and an update gate. The reset gate is used to determine the previous hidden state that will be forgotten, whereas the update gate determines how much of the new state is the old state and how much is a learned state. Bidirectional GRUs (Bi-GRU) are an extension of the basic GRU which allows the model to have both backward and forward information about the sequence at every time step. The input sequence is first fed in normal time order for the Forward GRU, and then in reverse time order for the Backward GRU. The outputs of the two GRUs are concatenated at each time step. Therefore, for each element in the sequence, Bi-GRU can access the previous information as well as the future information. The Bi-GRU unit can be explained by these equations \cite{grureview} \cite{bigrureview}:

Forward GRU:
\begin{itemize}
    \item Forward update gate: $z^f_t = sigmoid(W^f_z \cdot [h^f_{t-1}, x_t])$
    \item Forward reset gate: $r^f_t = sigmoid(W^f_r \cdot [h^f_{t-1}, x_t])$
    \item Forward candidate hidden state: $\tilde{h}^f_t = tanh(W^f \cdot [r^f_t \cdot h^f_{t-1}, x_t])$
    \item Forward final hidden state: $h^f_t = (1 - z^f_t) \cdot h^f_{t-1} + z^f_t \cdot \tilde{h}^f_t$
\end{itemize}

Backward GRU:
\begin{itemize}
    \item Backward update gate: $z^b_t = sigmoid(W^b_z \cdot [h^b_{t+1}, x_t])$
    \item Backward reset gate: $r^b_t = sigmoid(W^b_r \cdot [h^b_{t+1}, x_t])$
    \item Backward candidate hidden state: $\tilde{h}^b_t = tanh(W^b \cdot [r^b_t \cdot h^b_{t+1}, x_t])$
    \item Backward final hidden state: $h^b_t = (1 - z^b_t) \cdot h^b_{t+1} + z^b_t \cdot \tilde{h}^b_t$
\end{itemize}

where in these equations:
\begin{itemize}
    \item The superscript $f$ refers to parameters or states associated with the forward GRU, and the superscript $b$ refers to those associated with the backward GRU.
    \item $x_t$ is the input at time $t$
    \item $h^f_{t-1}$ and $h^b_{t+1}$ are the hidden states of the previous time step in the forward and backward GRUs, respectively
    \item $W^f_z$, $W^f_r$, $W^f$, $W^b_z$, $W^b_r$, $W^b$ are weight matrices learned during training for the forward and backward GRUs, respectively
    \item The sigmoid function squashes the output between 0 and 1, and the $tanh$ function squashes the output between -1 and 1
\end{itemize}

The Bi-GRU offers an advantage for music genre classification tasks. By processing data in both directions, it captures interdependencies between musical elements across time. This context is key in music, for example, the understanding of a musical note may depend on the notes that came before it as well as the notes after it. Therefore, Bi-GRU is expected to supplement the CNN architecture to provide a more comprehensive features for genre classification.

\subsection{Proposed Architecture}
\begin{figure}[h!]
    \centering 
    \includegraphics[scale=0.6]{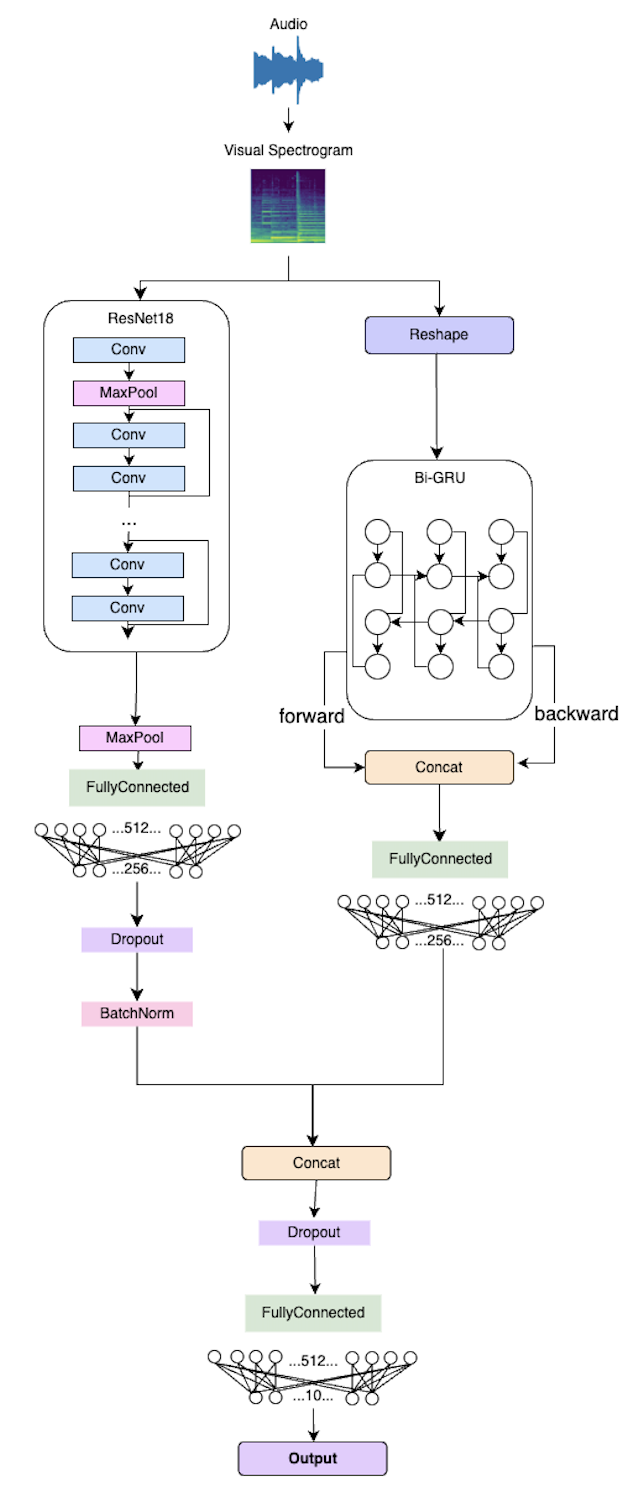}
    \caption{Parallel ResNet-BiGRU}
    \label{fig:proposed}
\end{figure}

The key idea of the proposed architecture is to combine two robust deep learning architectures, Residual Network (ResNet) and Bidirectional Gated Recurrent Unit (Bi-GRU), to provide a richer set of features for classification which can recognise both spatial hierarchical dependencies and temporal dependencies of the input. The architecture leverages the strengths of both ResNet18 and Bi-GRU, targeting two crucial aspects of the complex data structure. ResNet excels at extracting the intricate spatial hierarchical dependencies from input data which makes it excellent at handling visual spectrograms. Bi-GRU, on the other hand, capture temporal dependencies which is important with music data. By combining two structures, a musical piece can be studied with more comprehensive features for genre classification. The architecture combines ResNet and GRU in a parallel way where the input spectrogram is processed by two distinct pathways: a ResNet pathway and a Bi-GRU pathway before being concatenated and sent to a final fully connected classification layer. The model architecture is illustrated as Fig. \ref{fig:proposed}. A summary of the architecture is as follows: 

\begin{itemize}
    \item ResNet pathway
    \begin{itemize}
        \item ResNet18 (pre-trained) model (excluding the last layer)
        \item Adaptive Max Pooling Layer (1x1)
        \item Fully Connected Layer (512 input features, 256 output features)
        \item Dropout Layer (0.5 dropout rate)
        \item Batch Normalization Layer
    \end{itemize}
    \item Bi-GRU pathway
    \begin{itemize}
        \item Reshape
        \item GRU Layer (hidden size 256, 1 layer, bidirectional)
        \item Concatenation of forward output and backward output (512 features)
        \item Fully Connected Layer (512 input features, 256 output features)
    \end{itemize}
    \item Concatenation of ResNet and GRU outputs
    \item Dropout Layer (0.5 dropout rate)
    \item Fully Connected Layer (512 input features, classify into 10 genres)
\end{itemize}

For the ResNet pathway, the model uses a pre-trained ResNet18 model as the feature extractor to process the visual spectrograms. The last layer of the original ResNet18 is excluded, since the last layer classifies the input. This pathway begins by applying the ResNet model to the input data. The output is then passed through a max pooling 2-dimensional layer, which reduces the spatial dimensions to $1\times 1$ while preserving the number of feature maps. After the pooling, the output tensor is flattened and passed through a fully connected layer to reduce the number of features to 256. A dropout layer is applied for regularisation and a batch normalisation layer is applied for stabilising learning and reducing generalisation error. The ResNet pathway outputs a tensor with 256 features.

For the Bi-GRU pathway, the input dimensions are permuted to take the width of the image as a time sequence, where each column of pixels is treated as a timestep. The output is reshaped and then fed into a Bi-GRU. The pathway then takes the last step of the forward direction and the first step of the backward direction and concatenates them to form the final output, which has 256 features. 

The outputs of the ResNet and Bi-GRU pathways are concatenated along the feature dimension, resulting in 512 features. This is then passed through a fully connected layer to classify the input to 10 genres. The results of the classification will be analysed in the next section.

\section{Results}
\subsection{Metrics}
The metrics used to evaluate the performance is as follows:
\begin{itemize}
    \item Accuracy: accuracy refers to the total number of correct predictions divided by the total number of predictions. 
    \begin{equation}
        Accuracy = \frac{TP + TN}{TP + TN + FP + FN}
    \end{equation}
    \item Precision: precision is the total number of correct positive predictions divided by the total number of positive predictions. 
    \begin{equation}
        Precision = \frac{TP}{TP + FP}
    \end{equation}
    \item Recall: recall is the total number of correct positive predictions divided by the total number of actual positives. 
    \begin{equation}
        Recall = \frac{TP}{TP + FN}
    \end{equation}
    \item F1-score: F1 score is the harmonic mean of precision and recall. 
    \begin{equation}
        F1\textnormal{-}Score = 2 \cdot \frac{Precision \cdot Recall}{Precision + Recall}
    \end{equation}
\end{itemize}
The weighted average is employed given the context of multi-class classification, which calculates the above metrics for each class individually and then takes an average.

\subsection{Train/test Split}

To prepare the data into training and testing, a $DataLoader$ object is first created to load the images and labels. Subsequently, since the data is being augmented, $GroupShuffleSplit$ is implemented to prevent 'data leakage', which leads to overly optimistic performance during training but don't generalize well to unseen data (as shown by Table \ref{tab:performance2}). Rather than distributing the data across both the training and testing group, $GroupShuffleSplit$ segregate the images from the same song into either the training or testing group. This ensures a more robust attempt to train a model that learns general musical features, which gives a more realistic evaluation of how the model would perform in real world (Table \ref{tab:performance1}). The training set accounts for 80\% of the data whereas the testing set accounts for 20\% of the data. The reproducibility is maintained by setting a random see ($random\_state = 42$).

\begin{table}[h!]
    \centering
    \caption{Model Performance Without GroupShuffleSplit}
    \label{tab:performance2}
    \begin{tabular}{lcccccccc}
        \toprule
        Models & \multicolumn{4}{c}{STFT Spectrogram} & \multicolumn{4}{c}{Mel Spectrogram} \\
        \cmidrule(lr){2-5} \cmidrule(lr){6-9}
         & Precision & Recall & F1-score & Accuracy & Precision & Recall & F1-score & Accuracy\\
        \midrule
        CNN & 0.61 & 0.61 & 0.61 & 0.61 & 0.66 & 0.66 & 0.66 & 0.66 \\
        Bi-GRU & 0.71 & 0.71 & 0.71 & 0.71 & 0.78 & 0.78 & 0.78 & 0.78 \\
        ResNet18 & 0.70 & 0.70 & 0.70 & 0.70 & 0.87 & 0.87 & 0.87 & 0.87 \\
        ResNet-BiGRU & 0.86 & 0.86 & 0.86 & 0.86 & 0.90 & 0.90 & 0.90 & 0.90 \\
        \bottomrule
    \end{tabular}
\end{table}

\subsection{Results \& Analysis}
The training and testing process were conducted using early stopping. This study explores four architectures: CNN, Bi-GRU, ResNet18, and the hybrid model ResNet-BiGRU. Each model was trained and evaluated based on two types of spectrogram representations, namely, the STFT Spectrogram and Mel Spectrogram. The results are summarized in Table \ref{tab:performance1}.
\begin{table}[h!]
    \centering
    \caption{Model Performance Comparison With GroupShuffleSplit}
    \label{tab:performance1}
    \begin{tabular}{lcccccccc}
        \toprule
        Models & \multicolumn{4}{c}{STFT Spectrogram} & \multicolumn{4}{c}{Mel Spectrogram} \\
        \cmidrule(lr){2-5} \cmidrule(lr){6-9}
        & Precision & Recall & F1-score & Accuracy & Precision & Recall & F1-score & Accuracy \\
        \midrule
        CNN & 0.53 & 0.53 & 0.52 & 0.53 & 0.59 & 0.58 & 0.58 & 0.58 \\
        Bi-GRU & 0.50 & 0.50 & 0.49 & 0.50 & 0.55 & 0.55 & 0.54 & 0.55 \\
        ResNet18 & 0.55 & 0.55 & 0.54 & 0.55 & 0.79 & 0.79 & 0.78 & 0.79 \\
        ResNet-BiGRU & 0.75 & 0.75 & 0.75 & 0.75 & 0.82 & 0.81 & 0.81 & 0.81 \\
        \bottomrule
    \end{tabular}
\end{table}

Across all models, the results highlight a superior performance when utilizing Mel Spectrogram inputs, underscoring the advantage of this representation. Notably, the ResNet18 model demonstrated a substantial increase when moving from STFT Spectrogram to Mel Spectrogram. 
The highest performing model was the hybrid ResNet-BiGRU, achieving an accuracy of 0.75 for STFT Spectrogram inputs and 0.81 for the Mel Spectrogram inputs. For the STFT Spectrogram input, CNN, Bi-GRU, and ResNet18 showed comparable performance, indicating that the inherent differences among these architectures might not contribute to significant performance variations for STFT Spectrograms. Nevertheless, the hybrid model outperformed the others. This suggests that combining the strength of both convolutional and recurrent networks can have significant benefits by concurrently managing hierarchical and temporal dependencies within the data. On the other hand, for the Mel Spectrogram inputs, ResNet18 and the hybrid model performs similar and significantly outperforms the rest. This observation hints that Mel Spectrograms may provide a more effective representation of spatial data in the context of these models.

\begin{figure}[h!]
    \centering
    \begin{subfigure}{.33\textwidth}
        \centering
        \includegraphics[width=\linewidth]{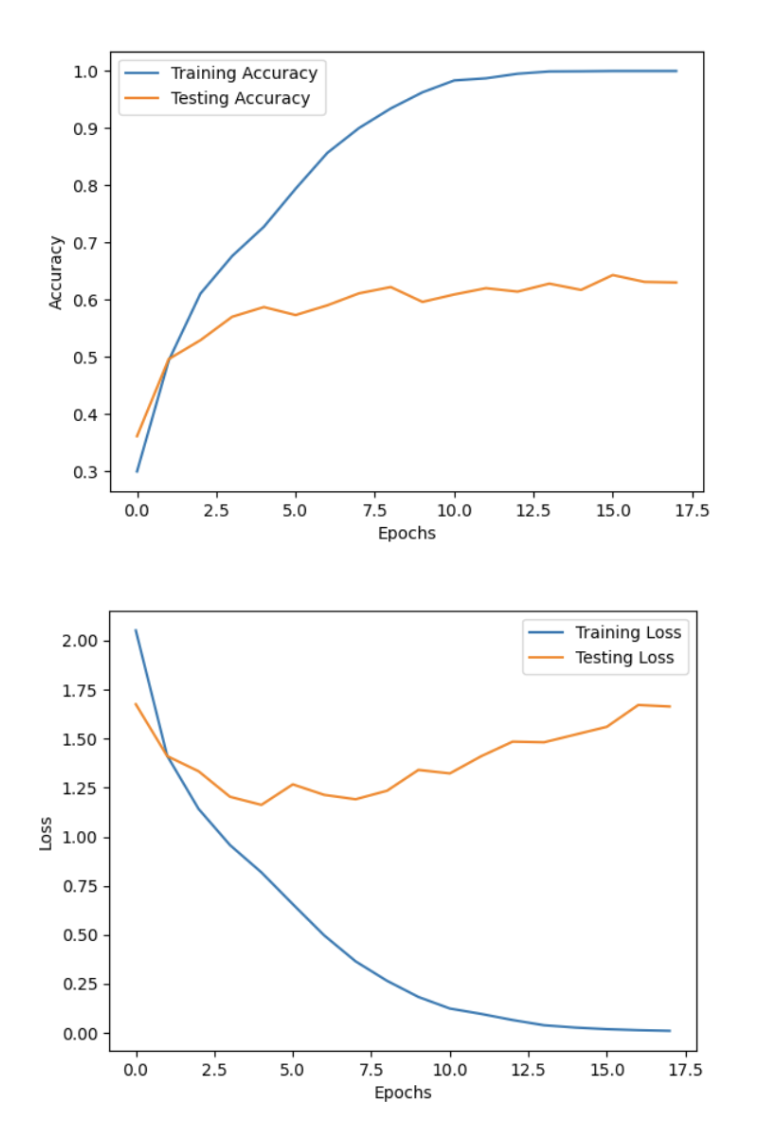}
        \caption{CNN}
        \label{fig:sub1}
    \end{subfigure}%
    \begin{subfigure}{.33\textwidth}
        \centering
        \includegraphics[width=\linewidth]{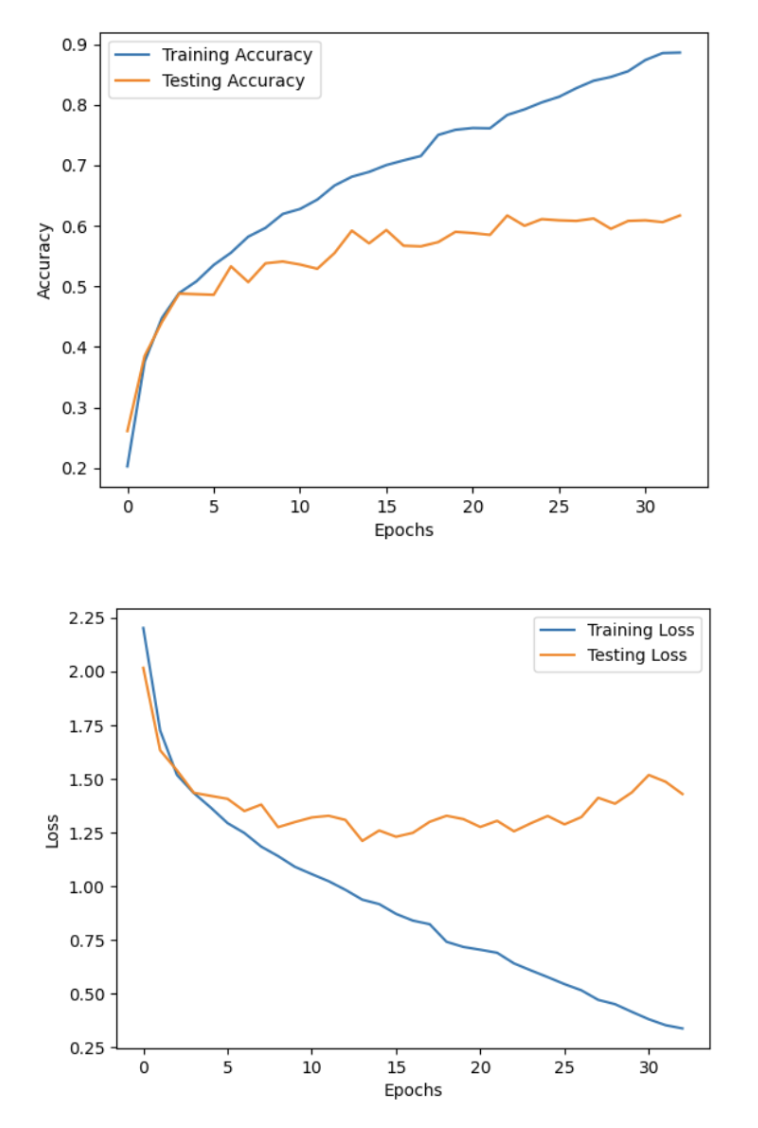}
        \caption{Bi-GRU}
        \label{fig:sub2}
    \end{subfigure}
    \begin{subfigure}{.33\textwidth}
        \centering
        \includegraphics[width=\linewidth]{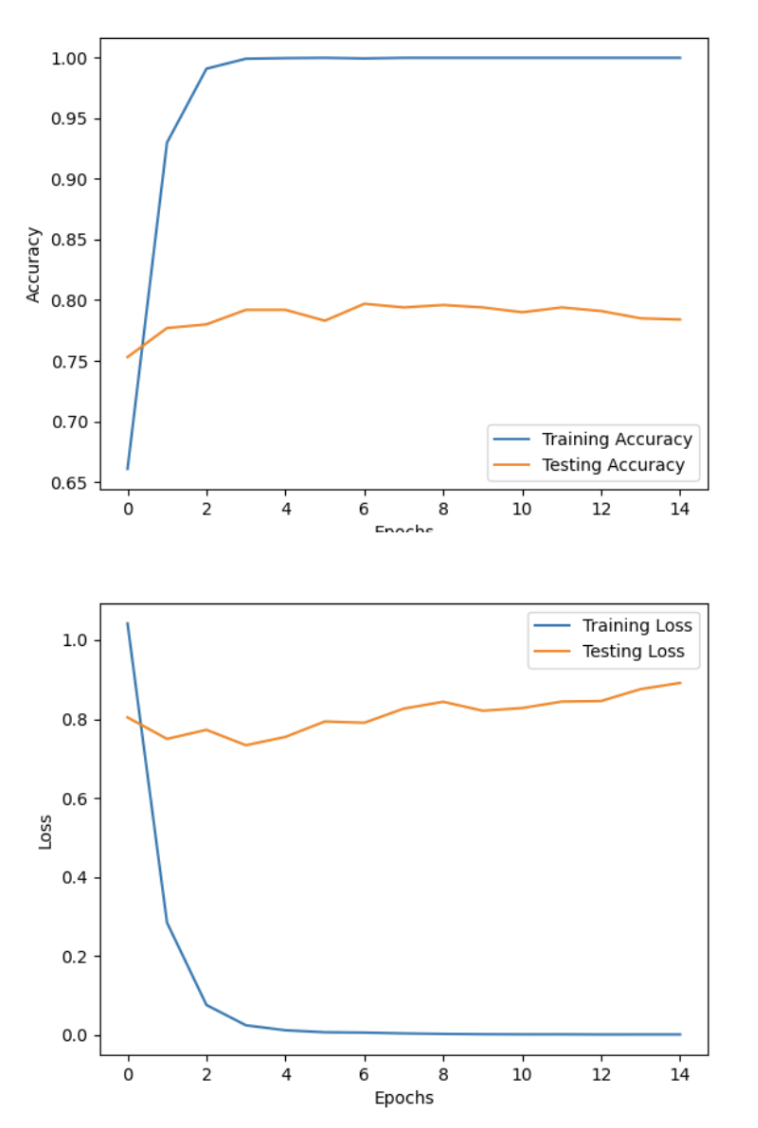}
        \caption{ResNet18}
        \label{fig:sub3}
    \end{subfigure}
    \caption{Accuracy and loss curves (input: Mel Spectrograms)}
    \label{fig:accloss}
\end{figure}

\begin{figure}[h!]
    \centering
    \begin{subfigure}{.45\textwidth}
        \centering
        \includegraphics[width=\linewidth]{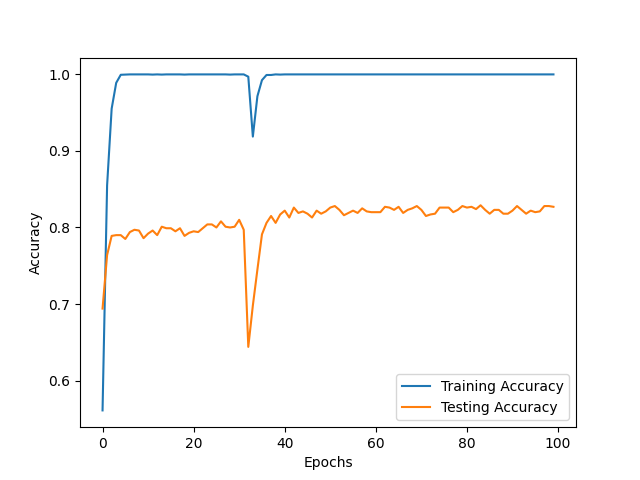}
        \caption{Accuracy curve}
        \label{fig:sub1}
    \end{subfigure}
    \begin{subfigure}{.45\textwidth}
        \centering
        \includegraphics[width=\linewidth]{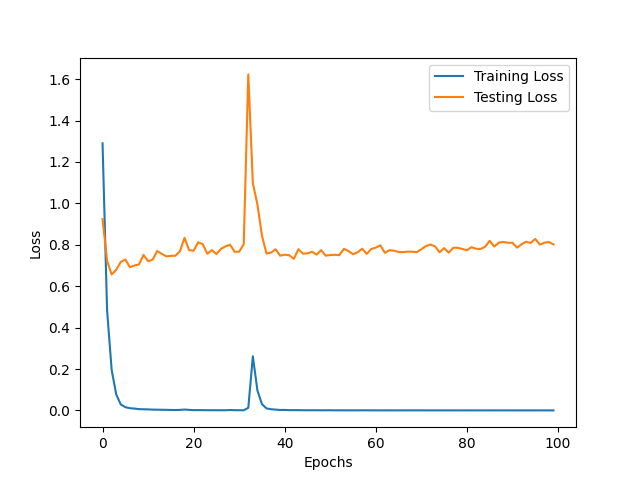}
        \caption{Loss curve}
        \label{fig:sub2}
    \end{subfigure}
    \caption{Accuracy and loss curves for ResNet-BiGRU (input: Mel Spectrograms)}
    \label{fig:acclosshybrid}
\end{figure}

To better analyse the model's learning patterns, the accuracy and loss across epochs is monitored. Fig. \ref{fig:accloss} and Fig. \ref{fig:acclosshybrid} illustrate the accuracy and loss curves for the four models when using Mel Spectrogram as input. All models implement early stopping with a patience setting of 10 epochs. From the results, it can be seen that all of the three baseline models (CNN, Bi-GRU, ResNet18) exhibit signs of overfitting, especially with the CNN and ResNet18. This could be due to the complexity of the models with the given amount of data. The Bi-GRU model appears to manage overfitting somewhat better. Therefore, for the hybrid model, a combination of batch normalisation and dropout layer is introduced for the ResNet path to deal with the overfitting issue. This addition allows the hybrid model to run through all 100 epochs instead of triggering early stopping. However, the accuracy plateaus at approximately 0.80. It is also worth noticing that the accuracy did not increase for more complex model using ResNet50, and increasing the number of layers for the Bi-GRU. This may indicate that the number of data is not large enough for the complexity of the model, or that the data itself may not be suitable for more complex architecture.

\begin{figure}[h!]
    \centering
    \begin{subfigure}{.35\textwidth}
        \centering
        \includegraphics[width=0.9\linewidth]{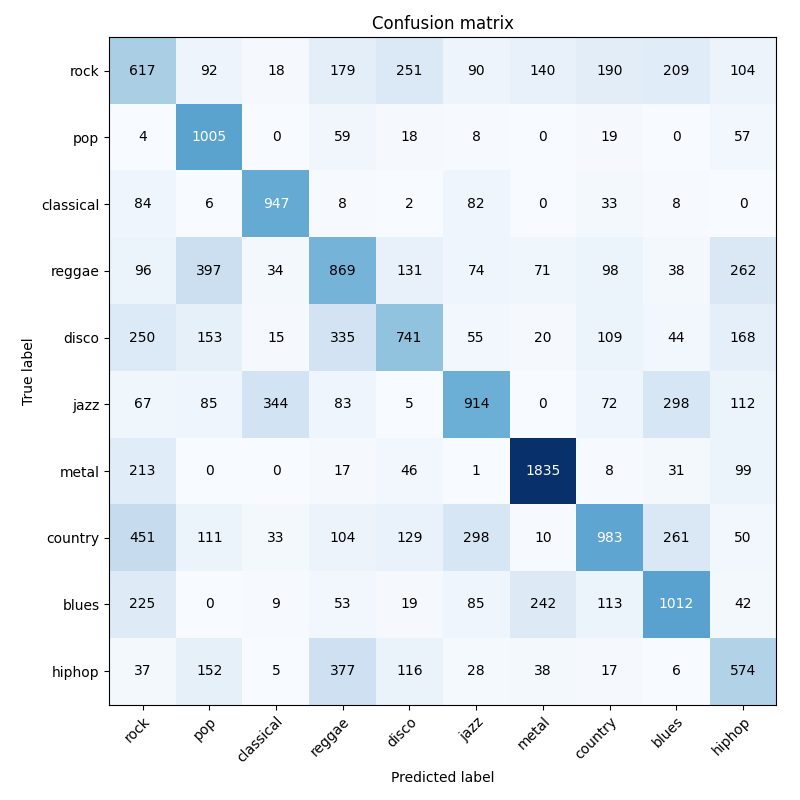}
        \caption{CNN}
        \label{fig:sub1}
    \end{subfigure}
    \begin{subfigure}{.35\textwidth}
        \centering
        \includegraphics[width=0.9\linewidth]{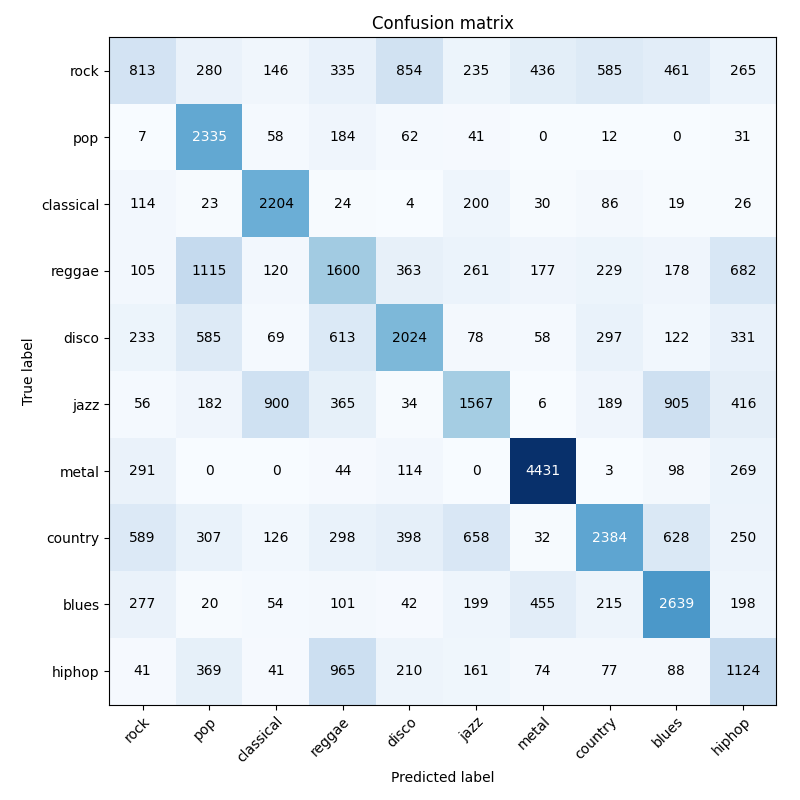}
        \caption{Bi-GRU}
        \label{fig:sub2}
    \end{subfigure}
    \begin{subfigure}{.35\textwidth}
        \centering
        \includegraphics[width=0.9\linewidth]{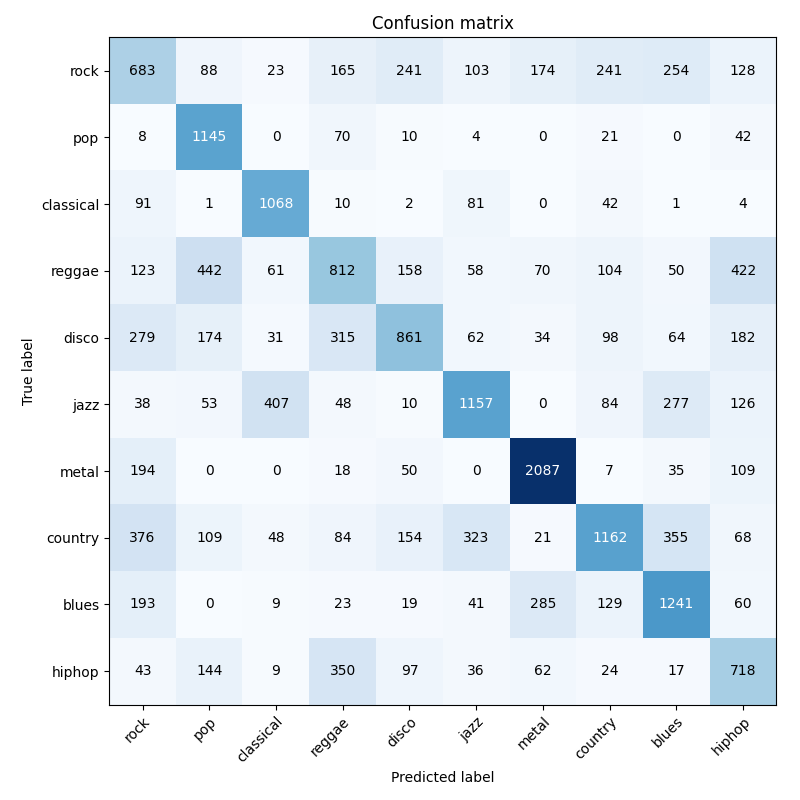}
        \caption{ResNet18}
        \label{fig:sub3}
    \end{subfigure}
    \begin{subfigure}{.35\textwidth}
        \centering
        \includegraphics[width=0.9\linewidth]{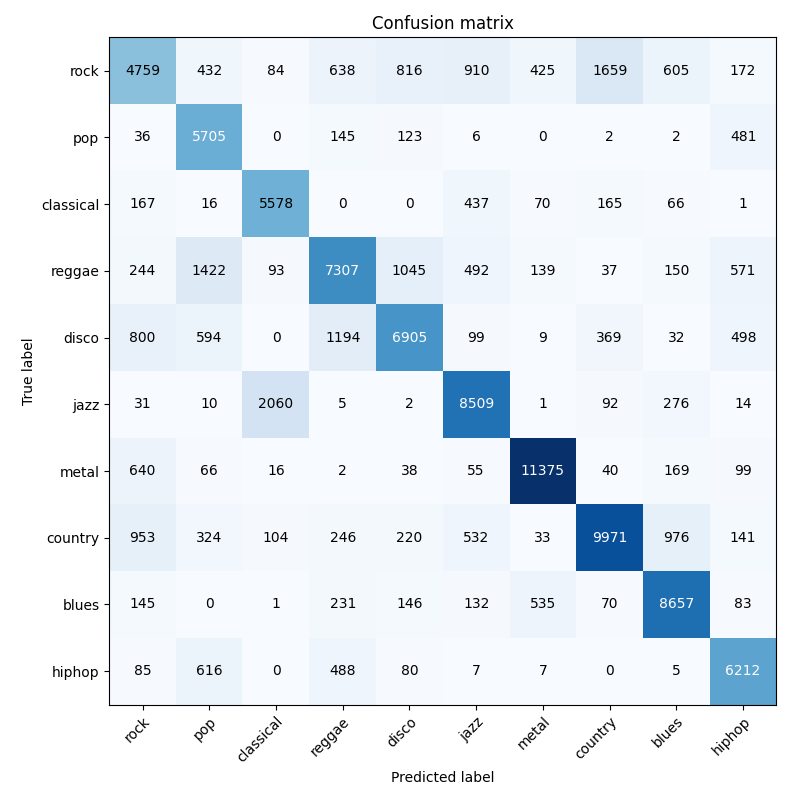}
        \caption{ResNet-BiGRU}
        \label{fig:sub4}
    \end{subfigure}
    \caption{Confusion Matrices (input: STFT Spectrograms)}
    \label{fig:comfusionmatrix}
\end{figure}

\begin{figure}[h!]
    \centering
    \begin{subfigure}{.35\textwidth}
        \centering
        \includegraphics[width=0.9\linewidth]{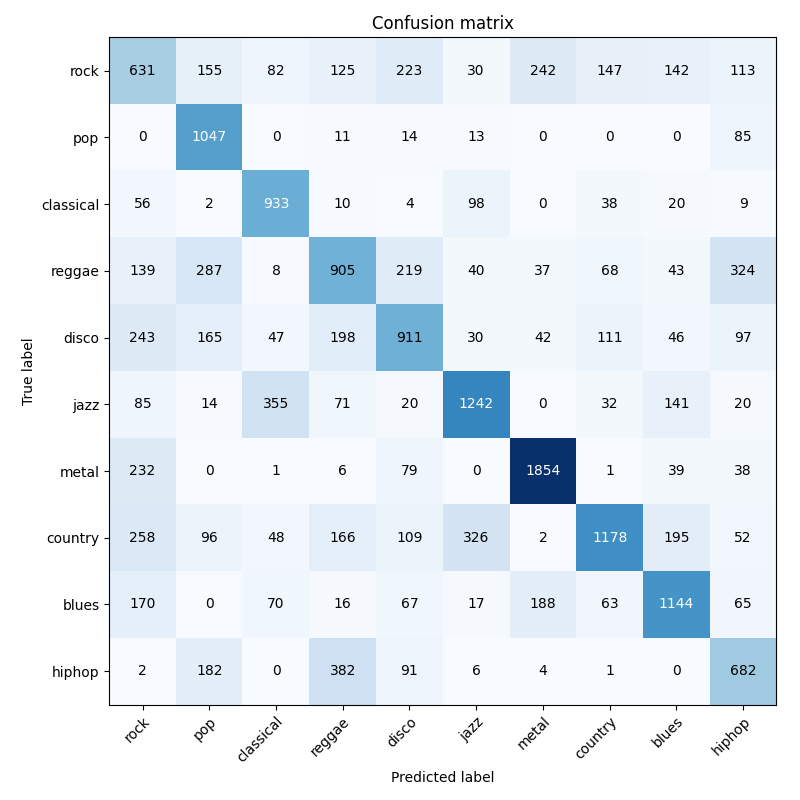}
        \caption{CNN}
        \label{fig:sub1}
    \end{subfigure}
    \begin{subfigure}{.35\textwidth}
        \centering
        \includegraphics[width=0.9\linewidth]{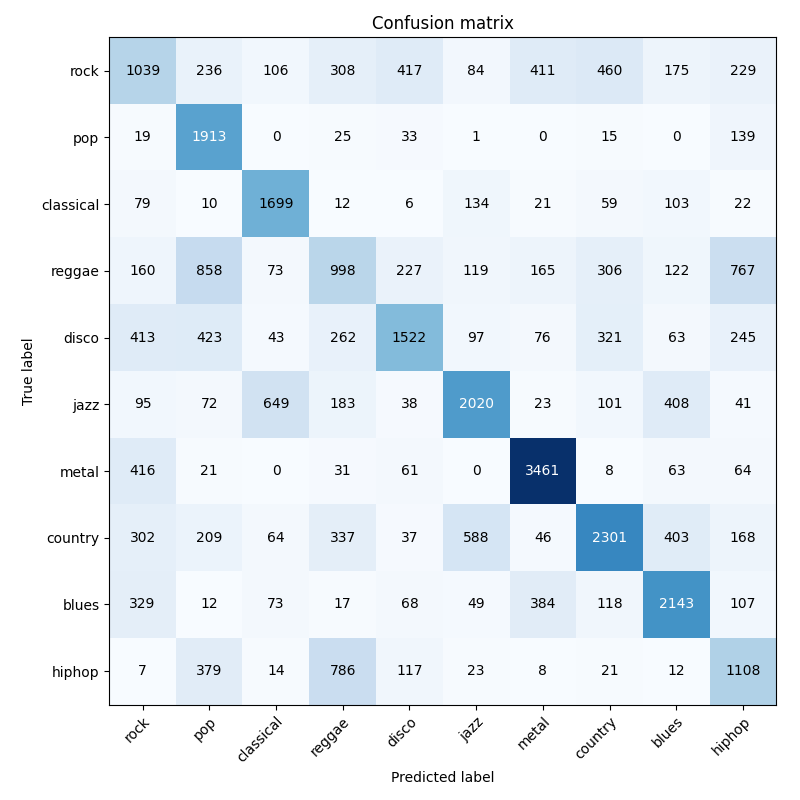}
        \caption{Bi-GRU}
        \label{fig:sub2}
    \end{subfigure}
    \begin{subfigure}{.35\textwidth}
        \centering
        \includegraphics[width=0.9\linewidth]{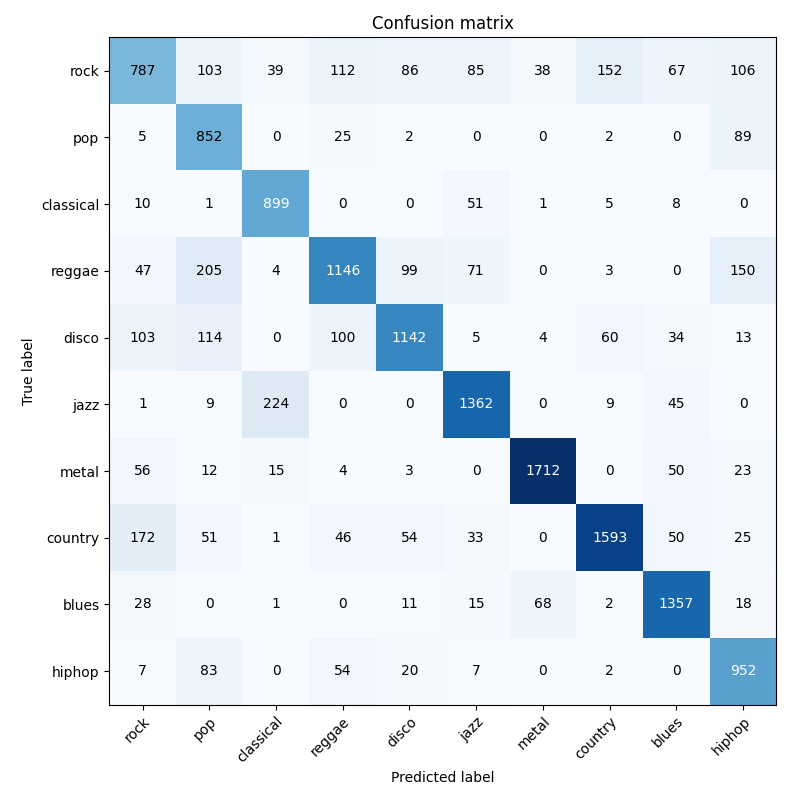}
        \caption{ResNet18}
        \label{fig:sub3}
    \end{subfigure}
    \begin{subfigure}{.35\textwidth}
        \centering
        \includegraphics[width=0.9\linewidth]{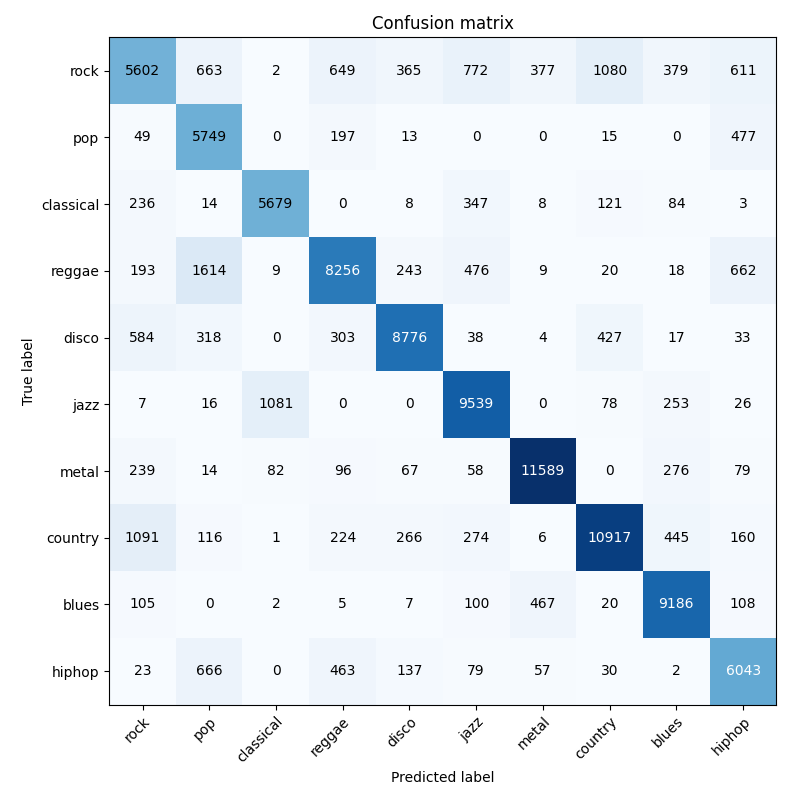}
        \caption{ResNet-BiGRU}
        \label{fig:sub4}
    \end{subfigure}
    \caption{Confusion Matrices (input: Mel Spectrograms)}
    \label{fig:comfusionmatrixmel}
\end{figure}

Fig. \ref{fig:comfusionmatrix} and Fig. \ref{fig:comfusionmatrixmel} display the confusion matrices for the two input representations. Overall, the model performs best when classifying the 'metal' genre and struggles the most with the 'rock' genre. This disparity may stem from the distinctive characteristics of these two genres. The 'metal' genre typically features highly amplified distortion, extended guitar solos, emphatic beats, and overall loudness \cite{metal}. These elements create unique patterns in the spectrograms, leading to better classification performance for this genre. In contrast, the 'rock' genre encompasses a range of styles, including rock and roll, electric blues, jazz music, country rock, even heavy metal\cite{wikirock}. This variety of sub-genres within 'rock' can have more varied patterns in the spectrograms, making it more challenging for the model to generalise. This might also explains the observation that a lot of the 'rock' class is being classified as 'disco' or 'country'.
The models also struggled to distinguish between 'reggae' and 'hiphop'. This confusion could arise from the overlapping musical elements and structure present in these genres. They share similar rhythmic components, often featuring a strong, syncopated bassline and rhythmic spoken or chanted lyrics\cite{wikireggae}. These commonalities could result in analogous patterns in the spectrograms, leading to classification errors. However, the ResNet-BiGRU hybrid model demonstrates better performance in distinguishing between 'reggae' and 'hiphop' genres when compared to the standalone CNN, Bi-GRU, and ResNet18 models. This could indicate that 'reggae' and 'hiphop', despite having overlapping elements, might exhibit subtle differences that the hybrid model can pick up.

\section{Discussion}
\subsection{Deep Learning Approach}
In this study, the application of deep learning models in the AMGC task has shown promising results when compared with the KNN and SVM baselines. The comparatively poor performance of the machine learning models (Table \ref{tab:accuracy2}) could be due to the absence of feature selection, such as using Principal Component Analysis or Random Projection. This highlights a key advantage of deep learning neural networks, which is their capability to automatically learn meaningful features from data, eliminating the need for manual feature engineering. This makes them more suitable for complex data, such as music data.
\subsection{Transfer Learning of Image Knowledge on Audio Data}
Another unique aspect of this study involves transferring knowledge from the image classification field to audio data analysis. By treating spectrograms as images, the deep learning architectures leveraged the power of established image classification networks (ResNet). As seen in the results, this approach is particularly effective. The finding underscores the potential of applying transfer learning to image and audio data.
\subsection{Hybrid Approach}
The hybrid ResNet-BiGRU model exhibited superior performance when using STFT Spectrograms as input compared to CNN, Bi-GRU, and ResNet. It also significantly outperforms CNN and Bi-GRU, and shows comparable performance with ResNet18 when Mel Spectrograms were used as input. This highlights the benefit of a hybrid approach which addresses a complex task by using a combination of architectures, each of which is suitable for a different aspect of the complexity inherent in the input data. This attests to the potential of dividing a complex task into parts and using different architectures to tackle different aspects of features inherent in the input data.
\subsection{Human Centered}
Another intriguing observation is that the difference in performance between using Mel Spectrograms and STFT Spectrograms depending on the input format and the underlying classification mechanism. As indicated by Table \ref{tab:accuracy2}, when using time series data as numerical inputs for machine learning models, the use of STFT Spectrogram as input features outperforms the use of Mel Spectrograms for the SVM model, while the KNN model is comparatively unaffected by these two representations. However, as shown by Table \ref{tab:performance1}, when image inputs are being used on deep learning models, Mel Spectrograms as input significantly outperform STFT Spectrograms. Since Mel Spectrograms are scaled using mel-scale, which better aligns with human auditory perception than the linearly-spaced frequency bands used in STFT Spectrograms \cite{melscale}, the deep learning approach using image input is more human centered. This supports the potential of integrating more human-aligned aspects into the AMGC task.

\section{Conclusion \& Future Work}
In conclusion, this study presents an innovative approach to the Automatic Music Genre Classification (AMGC) task by using visual spectrograms, and proposes a hybrid model that leverages the strengths of both Residual Networks (ResNet) and Gated Recurrent Units (GRU). The results reveal an accuracy of 0.81 when using visual Mel Spectrograms as input, significantly outperforming traditional machine learning models like KNN and SVM, and reaching the 70\% human classification benchmark. This underscores the suitability of deep learning for the AMGC task. The superior performance of the hybrid model compared to the others highlights its ability to better capture the spatial and temporal dynamics inherent in music data, showcasing the benefits of using a hybrid architecture to tackle different aspects of data complexity. The results also attest to the potential use of visual input in the AMGC task, underscoring the effectiveness of transfer learning from the field of image classification to audio data analysis. Further, the findings demonstrate that visual Mel Spectrogram input outperforms visual STFT Spectrogram input across all four deep learning architectures (CNN, Bi-GRU, ResNet18, ResNet-BiGRU), inspiring a more human-centered approach to the AMGC task.

Some future works can be inspired from this study:
\begin{itemize}
    \item Input representations: investigate other possible visual representations of audio data that resembles human auditory system (eg cochleagram) 
    \item Explore additional architectures: investigate other deep learning models to further attest to the benefit of using visual spectrograms as input
    \item Multitask learning: the model could be extended to perform other tasks such as music mood detection, or could be transformed into multi-purpose music analysis tool
    \item Generative model: the learned features from the model can be used to perform generative tasks such as genre shift
    \item Interpretability: using techniques to interpret the neural network's decisions is beneficial in improving the model, as well as reveal novel insights about music genres.
\end{itemize}

\section*{Acknowledgments}
This research was conducted as part of the SCIE30001 course at The University of Melbourne. I am grateful for the opportunity to undertake this project and for the resources provided by the school, which were instrumental in the completion of this research paper.

I would also like to express my deepest appreciation to my supervisor, Prof. Lars Kulik, and Nestor Cabello. Their invaluable guidance, continuous support, and constructive feedback have been instrumental in shaping this research. 

\newpage
\appendices
\section{Deployment of a music recommender using deep learning classification}
\begin{figure}[h!]
    \centering 
    \includegraphics[scale=1]{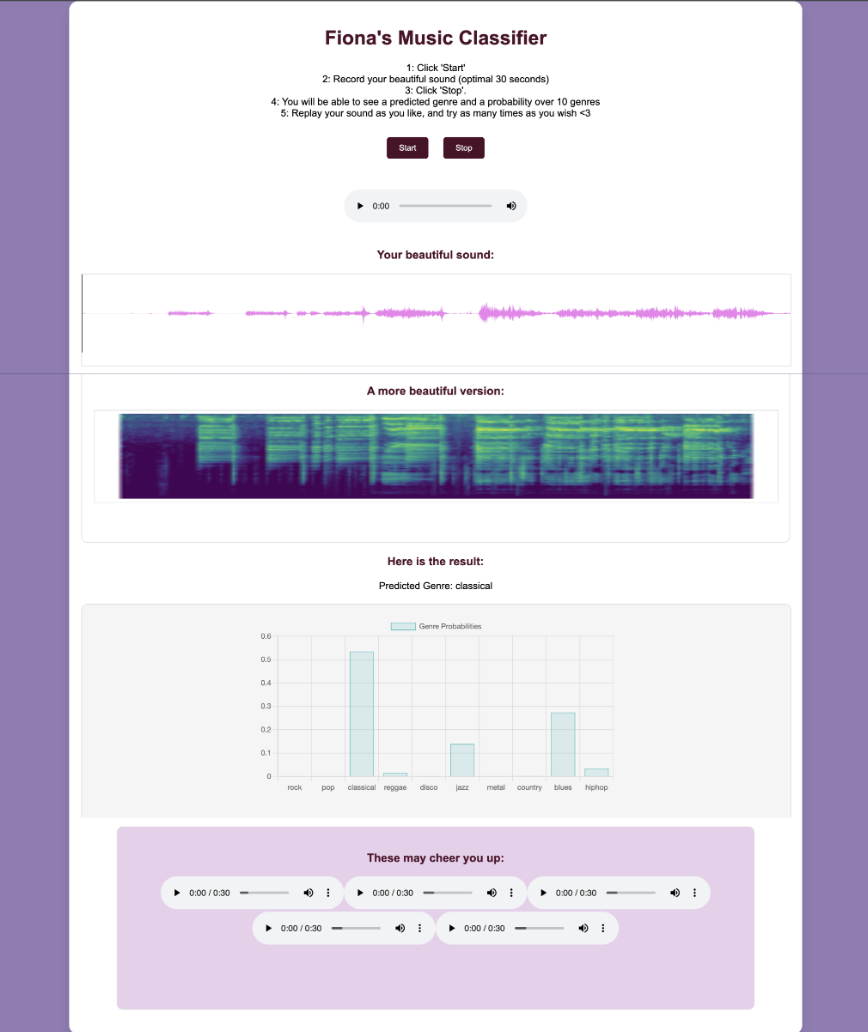}
    \caption{Deployment}
    \label{fig:muse}
\end{figure}

An intriguing product of this project is a small-scale music recommendation system WebApp, available at \url{https://deeplearnmuse-3t5mgrwzwa-km.a.run.app/}. The WebApp prompts the user to provide an audio input, which is then converted into a visual Mel Spectrogram and fed into a pretrained deep learning model for classification. The classification result, as well as the probabilities for the 10 genres, are returned to the user. Subsequently, five songs that have the most similar class distribution to the predicted probability distribution are recommended to the user. A screenshot of the WebApp is shown in Fig \ref{fig:muse}.

\end{document}